# Linear response strength functions with iterative Arnoldi diagonalization

J. Toivanen,[1] B.G. Carlsson,[1] J. Dobaczewski,[1,2] K. Mizuyama,[1]
R.R. Rodríguez-Guzmán,[1] P. Toivanen,[1] and P. Veselý[1]

[1]*Department of Physics, University of Jyväskylä, P.O. box 35, FIN-40014, Finland*
[2]*Institute of Theoretical Physics, Warsaw University, ul. Hoża 69, PL-00681, Warsaw, Poland*
(Dated: November 7, 2018)

We report on an implementation of a new method to calculate RPA strength functions with iterative non-hermitian Arnoldi diagonalization method, which does not explicitly calculate and store the RPA matrix. We discuss the treatment of spurious modes, numerical stability, and how the method scales as the used model space is enlarged. We perform the particle-hole RPA benchmark calculations for double magic nucleus $^{132}$Sn and compare the resulting electromagnetic strength functions against those obtained within the standard RPA.

## I. INTRODUCTION

The linear response theory (LRT) obtained from the linearized time-dependent mean field method is an important tool for calculating properties of excited states of many-fermion systems, such as nuclear giant resonances. In its charge-changing version, it can also give access to the beta-decay strengths. This method is especially important in heavy nuclei, where the shell-model or configuration-interaction approaches are intractable. An advantage is also that LRT does not require the knowledge of an interaction and can therefore be used both within density functional theory (DFT) and phenomenological energy density functional (EDF) approaches, giving rise to a set of equations of RPA type. Below, for simplicity we refer to this method and associated equations simply as RPA method/equations. Strength functions obtained in this way probe new aspects of the EDFs and thus have a potential of constraining parameters in phenomenological nuclear EDFs.

The purpose of the present study is to present an implementation of an efficient RPA algorithm that is based on the local nuclear EDF. For electronic systems, similar methods have been used since many years, see, e.g., the recent Ref. [1] for a review, and they also constitute parts of standardized computer packages such as GAMESS [2, 3]. There are two essential elements of these methods, which are at the heart of their efficiency and scalability, namely, (i) the RPA equations are solved iteratively and (ii) the RPA matrix does not have to be explicitly calculated. The second of these elements is particularly important; it is based on the observation that the action of the RPA matrix on the vector of RPA amplitudes can proceed through the calculation of the mean fields corresponding to these amplitudes.

In nuclear physics context, probably the first study that used the concept of mean fields in the RPA method was that by P.-G. Reinhard [4]. Iterative solutions of the RPA equations were introduced by Johnson, Bertsch, and Hazelton [5], and applied to the case of separable interactions, but in fact these methods can also be applied in more complicated situations, as we show here. Strangely enough, these very efficient methods have not yet been used in practical applications. Only very recently, Nakatsukasa *et al.* [6, 7] have implemented the analogous approach within the so-called finite amplitude method (FAM).

Our present implementation pertains to the spherical symmetry with neglected pairing correlations – thus it constitutes only a proof-of-principle study. The real challenge is in solving the quasiparticle RPA (QRPA) problem in deformed nuclei. Although at present, a few implementations that are based on solving the standard QRPA equations already exist [8, 9] or begin to emerge [10, 11], such a route is bound to be blocked by the shear dimensionality of the problem. On the other hand, as we show here, methods based on the iterative solutions using mean fields have much better scalability properties and are potentially very promising.

The paper is organized as follows. In Secs. II and III we lay down the essential features of the method by presenting the use of mean fields and iterative solution of the RPA method, respectively. Then, in Sec. IV we present the method to remove the spurious RPA states, which is tailored to be used within the iterative approach. Sections V and VI present the convergence and scalability properties of our method, respectively, and summary and conclusions are given in Sec. VII.

## II. RPA FROM LINEARIZED TDHF

To be concise, below we present a less general derivation than the standard method [12, 13] to derive the RPA equations from linearized time-dependent Hartree-Fock (TDHF) equations. For density independent forces or functionals with terms quadratic in density, the density matrix and mean field of a time-dependent nuclear state are expressed as

$$\rho(t) = \rho_0 + \tilde{\rho}_\omega e^{i\omega t} + \tilde{\rho}_\omega^\dagger e^{-i\omega t}, \qquad (1)$$

$$h(t) = h_0 + \tilde{h}_\omega e^{i\omega t} + \tilde{h}_\omega^\dagger e^{-i\omega t}, \quad (2)$$

where $\rho_0$ and $h_0$ are the Hartree-Fock (HF) ground-state density matrix and mean field, respectively. Inserting $\rho(t)$ and $h(t)$ of Eqs. (1) and (2) into the TDHF equation, and keeping only terms linear in the fluctuating quantities $\tilde{\rho}$ and $\tilde{h}$, we get a linearized TDHF equation, or the RPA equations:

$$\hbar\omega\tilde{\rho}_{\omega,mi} = (\epsilon_m - \epsilon_i)\tilde{\rho}_{\omega,mi} + \tilde{h}_{\omega,mi}, \quad (3)$$
$$\hbar\omega\tilde{\rho}_{\omega,im} = (\epsilon_i - \epsilon_m)\tilde{\rho}_{\omega,im} - \tilde{h}_{\omega,im}, \quad (4)$$

where we use the letter $m$ for particle states and $i$ for hole states, and where $\epsilon_{m,i}$ are the HF single-particle energies. The fields $\tilde{h}$ are the first functional derivatives of the used EDF, evaluated using the density amplitudes $\tilde{\rho}$ of Eq. (1). Density dependence of the used EDF beyond quadratic gives rise to rearrangement fields in $\tilde{h}$. These rearrangement parts of $\tilde{h}$ must be linearized around $\rho_0$ to make our RPA equations explicitly first order in $\tilde{\rho}$.

One way to achieve this is by calculating functional derivatives of the rearrangement parts of $\tilde{h}$ with respect to density, which technically makes our mean-field routine differ from the standard HF routines. Since in our implementation we use the standard Skyrme forces that have simple density dependencies of the coupling constants, the explicit functional differentiation does not cause any mathematical or performance problems. Had a EDF with more complex density dependence been used it would have been an advantage to instead use the FAM method [6] for linearization.

If the matrix elements of $\tilde{h}$ in Eqs. (3) and (4) are expanded in terms of the particle-hole (p-h) and hole-particle (h-p) matrix elements of $\tilde{\rho}$, we obtain the traditional RPA equations. In this work, we do not construct the RPA matrix, but directly solve Eqs. (3) and (4) by calculating the matrix elements of fields $\tilde{h}$ using a HF mean-field routine that uses the time-reversal-invariance breaking density matrix $\tilde{\rho}$. Since the same routine is used to evaluate the HF and RPA mean fields, the method is always fully self-consistent [14, 15]. In the following equations, we use the standard abbreviations $X_{mi} = \tilde{\rho}_{\omega,mi}$ and $Y_{mi} = \tilde{\rho}_{\omega,im}$. The density vector that contains the p-h matrix elements of $\tilde{\rho}^\omega$ is defined as $\mathcal{X}^\omega = (X_{m_1,i_1}, X_{m_2,i_2}, \ldots, X_{m_D,i_D})$, and similarly for the vector $\mathcal{Y}$ of h-p elements, where $D$ is the number of allowed p-h configurations. Overlaps of RPA vectors are defined as

$$\langle X, Y | X', Y' \rangle = (\mathcal{X}^*, \mathcal{Y}^*) \begin{pmatrix} \mathcal{X}'^T \\ -\mathcal{Y}'^T \end{pmatrix} \quad (5)$$

and the minus sign results from the RPA norm matrix.

### III. ITERATIVE SOLUTION OF THE RPA EQUATIONS

The RPA equations (3) and (4) constitute a non-hermitian eigenproblem with non-positive-definite norm. We solve this problem by using an iterative method that during each iteration only needs to know the product of the RPA matrix and a density vector, that is, the right-hand sides of Eqs. (3) and (4):

$$W_{mi}^k = (\epsilon_m - \epsilon_i)X_{mi}^k + \tilde{h}_{mi}(\mathcal{X}^k, \mathcal{Y}^k), \quad (6)$$
$$W'^k_{mi} = (\epsilon_i - \epsilon_m)Y_{mi}^k - \tilde{h}_{im}(\mathcal{X}^k, \mathcal{Y}^k), \quad (7)$$

where index $k$ labels iterations and the mean fields $\tilde{h}(\mathcal{X}^k, \mathcal{Y}^k)$ depend linearly on the density vectors $\mathcal{X}^k$ and $\mathcal{Y}^k$. Expressed through the standard RPA matrices $A$ and $B$ [12], Eqs. (6) and (7) for a positive norm basis vector and for its opposite norm partner vector read:

$$\begin{pmatrix} \mathcal{W}_+^k \\ \mathcal{W}_+'^k \end{pmatrix} = \begin{pmatrix} A & B \\ -B'^* & -A'^* \end{pmatrix} \begin{pmatrix} \mathcal{X}^k \\ \mathcal{Y}^k \end{pmatrix}, \quad (8)$$

$$\begin{pmatrix} \mathcal{W}_-^k \\ \mathcal{W}_-'^k \end{pmatrix} = \begin{pmatrix} A & B \\ -B'^* & -A'^* \end{pmatrix} \begin{pmatrix} \mathcal{Y}^{k*} \\ \mathcal{X}^{k*} \end{pmatrix}. \quad (9)$$

In exact arithmetic $A = A'$ and $B = B'$ and therefore either Eqs. (8) or (9) could be used in the iteration procedure with equivalent results. Nevertheless, below we use them both to stabilize the iteration process.

Various iterative methods, which only need to know the products of the diagonalized matrix and vectors, exist for non-hermitian matrix eigenvalue equations, and good examples with pseudocode are shown in Ref. [1]. We chose the non-hermitian Lanczos method [5] in a modified form, because it conserves all odd-power energy weighed sum rules (EWSR) if the starting vector (pivot) of iteration is chosen correctly. In this work, we start from a pivot vector that has its elements set to the matrix elements of electromagnetic multipole operator,

$$X_{mi}^1 = \frac{e}{\sqrt{N^1}} \langle \phi_m | r^p Y_{JM} | \phi_i \rangle, \quad Y_{mi}^1 = 0, \quad (10)$$

where $p = 2$ and $J = 0$ for the $0^+$ mode, $p = 1$ and $J = 1$ for the $1^-$ mode, and $p = 2$ and $J = 2$ for the $2^+$ mode. The constant $N^1$ is used to normalize the pivot vector to unity. For the IS $1^-$ mode we only present results obtained with the operator:

$$(r^3 - \tfrac{5}{3}\langle r^2 \rangle r) Y_{1M}, \quad (11)$$

to stay consistent with Refs. [16] and [17]. For the choice of pivot in (10) one can prove [5] that all odd-power EWSRs are satisfied throughout the iteration procedure.

Because we calculate the RPA matrix-vector products by using the mean-field method, and not with a precalculated RPA matrix, we introduce small, but significant numerical noise to the resulting vectors. If corrective measures are not used to remove or reduce this noise, the iteration method fails and produces complex RPA eigenvalues early on in the iteration. We stabilize our iterative solution method by modifying the method of Ref. [5] in two ways. First, we use the non-hermitian Arnoldi method instead of the non-hermitian Lanczos method. The advantage of Arnoldi method is that it orthogonalizes each new basis vector against all previous basis vectors and their opposite norm partners, that is,

$$\begin{pmatrix} \tilde{\mathcal{X}}^{k+1} \\ \tilde{\mathcal{Y}}^{k+1} \end{pmatrix} = \begin{pmatrix} \mathcal{W}_+^k \\ \mathcal{W}'^{\,k}_+ \end{pmatrix} - \sum_{i=1}^{k} \begin{pmatrix} \mathcal{X}^i \\ \mathcal{Y}^i \end{pmatrix} a_{ik} + \sum_{i=1}^{k} \begin{pmatrix} \mathcal{Y}^{i*} \\ \mathcal{X}^{i*} \end{pmatrix} b_{ik}, \quad (12)$$

$$\begin{pmatrix} \tilde{\mathcal{Y}}^{k+1*} \\ \tilde{\mathcal{X}}^{k+1*} \end{pmatrix} = -\begin{pmatrix} \mathcal{W}_-^k \\ \mathcal{W}'^{\,k}_- \end{pmatrix} + \sum_{i=1}^{k} \begin{pmatrix} \mathcal{X}^i \\ \mathcal{Y}^i \end{pmatrix} b'^{*}_{ik} - \sum_{i=1}^{k} \begin{pmatrix} \mathcal{Y}^{i*} \\ \mathcal{X}^{i*} \end{pmatrix} a'^{*}_{ik}, \quad (13)$$

where the overlap matrices $a_{ik}$, $b_{ik}$, $a'^{*}_{ik}$, and $b'^{*}_{ik}$ are calculated as in Eq. (5). Again, in exact arithmetic, Eqs. (12) and (13) are equivalent and the lower matrix elements $a'^{*}_{ik}$ and $b'^{*}_{ik}$ in Eq. (13) are exact complex conjugates of the elements of the upper Krylov-space [18] RPA matrices. We assume this to keep the Krylov-space RPA matrix in the standard form.

In the Lanczos method, only the tridiagonal parts of RPA matrices are calculated, and small changes in basis vectors due to Lanczos re-orthogonalization (which always must be used to preserve orthogonality) do not show up in the constructed RPA matrix. In the Arnoldi method, these small but important elements outside the tridignal part improve the stability as compared to Lanczos.

The norm of the obtained new residual vector in Eq. (12) can be either positive or negative. We do not in practice use Eq. (13), which in exact arithmetic would duplicate the results of Eq. (12). Instead, we store only the positive-norm basis states and use a similar method as in Ref. [5] to change sign of the norm in case the norm of the residual vector in Eq. (12) is negative. Thus, explicitly, for the positive norm of the residual vector $\tilde{N}^{k+1} = \langle \tilde{X}^{k+1}, \tilde{Y}^{k+1} | \tilde{X}^{k+1}, \tilde{Y}^{k+1} \rangle$, we define the new normalized positive-norm basis vector as

$$X^{k+1}_{mi} = \frac{1}{\sqrt{\tilde{N}^{k+1}}} \tilde{X}^{k+1}_{mi}, \quad Y^{k+1}_{mi} = \frac{1}{\sqrt{\tilde{N}^{k+1}}} \tilde{Y}^{k+1}_{mi}. \quad (14)$$

If $\tilde{N}^{k+1} < 0$, the new normalized positive norm basis vector is defined as

$$X^{k+1}_{mi} = \frac{1}{\sqrt{-\tilde{N}^{k+1}}} \tilde{Y}^{k+1*}_{mi}, \quad Y^{k+1}_{mi} = \frac{1}{\sqrt{-\tilde{N}^{k+1}}} \tilde{X}^{k+1*}_{mi}. \quad (15)$$

When maximum number of iterations has been made or the iteration has been stopped, the generated Krylov-space RPA matrix, with dimension $d << D$, is diagonalized with standard methods, that is we solve:

$$\begin{pmatrix} a & b \\ -b^* & -a^* \end{pmatrix} \begin{pmatrix} x^k \\ y^k \end{pmatrix} = \hbar \omega_k \begin{pmatrix} x^k \\ y^k \end{pmatrix}. \quad (16)$$

The approximate RPA solutions are then obtained by transforming the Krylov-space basis vectors,

$$X^k_{mi} = \sum_{l=1}^{d} \left( X^l_{mi} x^k_l + Y^{l*}_{mi} y^{k*}_l \right), \quad (17)$$

$$Y^k_{mi} = \sum_{l=1}^{d} \left( Y^l_{mi} x^{k*}_l + X^{l*}_{mi} y^k_l \right), \quad (18)$$

for all $k = 1, \ldots, d$, and these vectors are used to evaluate the strength functions.

Standard RPA method that constructs and diagonalizes the full RPA matrix can ensure that the lower matrices in the RPA supermatrix are exact complex conjugates of the upper matrices. Our mean-field method can have small differences in the implicitly used upper and lower RPA matrices due to finite numerical precision. The consequence of this is that we will in general have $a'_{ij} \neq a_{ij}$ and $b'_{ij} \neq b_{ij}$. This spoils the consistency of Eqs. (12) and (13) and can make the Arnoldi iteration to fail and produce complex energy solutions.

The numerical errors in the matrix-vector products can be reduced by symmetrization. We thus calculate the RPA fields twice, first using the densities of a positive norm basis vector $(\mathcal{X}^k, \mathcal{Y}^k)$, and second using the densities of negative norm vector $(\mathcal{Y}^{k*}, \mathcal{X}^{k*})$. The two resulting vectors are subtracted from each other to get the final stabilized RPA matrix-vector product,

$$\begin{pmatrix} \mathcal{W}^k \\ \mathcal{W}'^k \end{pmatrix} = \frac{1}{2} \begin{pmatrix} \mathcal{W}_+^k - \mathcal{W}'^{\,k*}_- \\ \mathcal{W}'^{\,k}_+ - \mathcal{W}^{k*}_- \end{pmatrix}, \quad (19)$$

to be used in Eq. (12). Together, the Arnoldi iteration method and symmetrization of matrix-vector products stabilize our mean-field based iterative diagonalization.

## IV. TREATMENT OF SPURIOUS RPA MODES

For the discussion of various spurious modes in the RPA method we refer the reader to, e.g., Ref. [13]. In the present study, we only consider spherical ground states neglecting pairing correlations, so the only spurious excitation is generated by the total linear momentum. Therefore, the only affected RPA mode is the isoscalar $1^-$ mode. In traditional RPA calculations that construct and diagonalize the full RPA matrix, the spurious $1^-$ mode is typically removed *after* the RPA diagonalization. Often a modified transition operator (11) is used, which has the property of $\langle HF | \left[ \hat{F}, \hat{P}_{cm} \right] | HF \rangle = 0$, as long as the commutator is evaluated within a complete set of basis states. In a finite model spaces of localized orbitals this relation is no more exactly valid, and the corrected operator does not remove spurious components exactly.

To remove the spurious isoscalar $1^-$ mode from our physical RPA excitations we use the same method as in Ref. [6], where the basis vectors are orthogonalized against the spurious translational mode **P** and its conjugate "boost" operator **R**, which have the form:

$$\hat{P}_\mu = \frac{1}{\sqrt{3}} \sum_{mi} \left( i(\phi_m || \nabla_1 || \phi_i) \left[ c^\dagger_m \tilde{c}_i \right]_{1\mu} + \text{h.c.} \right), \quad (20)$$

$$\hat{R}_\mu = \frac{1}{\sqrt{3}} \sum_{mi} \left( (\phi_m || r_1 || \phi_i) \left[ c^\dagger_m \tilde{c}_i \right]_{1\mu} + \text{h.c.} \right). \quad (21)$$

The spurious RPA vectors $(\mathcal{P},\mathcal{P}^*)$ and $(\mathcal{R},\mathcal{R}^*)$ contain the p-h and h-p matrix elements of Eqs. (20) and (21), respectively. Our method differs from that of Ref. [6] in the fact that we orthogonalize our basis *during* the Arnoldi iteration, which fits naturally with the iterative solution method and guarantees that the obtained approximate RPA excitations have exact zero overlaps with spurious modes. This is equivalent to diagonalizing the full RPA matrix in the subspace orthogonal to the spurious states. In our implementation, each generated new Arnoldi basis vector is orthogonalized as

$$\begin{pmatrix} \mathcal{X}_k \\ \mathcal{Y}_k \end{pmatrix}_{phys.} = \begin{pmatrix} \mathcal{X}_k \\ \mathcal{Y}_k \end{pmatrix} - \lambda \begin{pmatrix} \mathcal{P} \\ \mathcal{P}^* \end{pmatrix} - \mu \begin{pmatrix} \mathcal{R} \\ \mathcal{R}^* \end{pmatrix}, \quad (22)$$

where the overlaps $\lambda$ and $\mu$ are defined as

$$\lambda = \frac{\langle R, R^* | X^k, Y^k \rangle}{\langle R, R^* | P, P^* \rangle}, \quad (23)$$

$$\mu = -\frac{\langle P, P^* | X^k, Y^k \rangle}{\langle R, R^* | P, P^* \rangle}. \quad (24)$$

When more symmetries are broken, formulas equivalent to Eqs. (22)–(24) can be used to remove spurious components coming from each broken symmetry of the mean field.

## V. CONVERGENCE OF STRENGTH FUNCTIONS

The iterative Arnoldi method is meaningful for the calculation of strength functions only if the number of iterations needed for accurate results is significantly less than the full RPA dimension. To study how many Arnoldi iterations we need for good accuracy, we calculated electromagnetic isoscalar (IS) and isovector (IV) strength functions [16] for doubly magic nuclei. All calculations were performed by implementing the RPA iterative solutions within the computer program HOSPHE [19], which solves the self-consistent equations in the spherical harmonic-oscillator (HO) basis. We studied both the convergence of smoothed strength functions as a function of number of Arnoldi iterations and as a function of the number of HO shells.

We used the same definitions of the $0^+$, $1^-$, and $2^+$ transition operators as in Ref. [17] and the Skyrme functional SLy4 of Ref. [20]. The function we used to smooth the strength functions was also the same as in [16], with $R_{box} = 20$ fm. Because the HF ground state of $^{132}$Sn is spherically symmetric, our approximate RPA phonons have good angular momentum. We tested the use of large basis sets up to 40 HO shells. The HF ground state energies were well converged for all double magic nuclei when 25 HO shells were used. Below, we present the results only for $^{132}$Sn.

### A. Convergence as a function of the number of Arnoldi iterations

#### 1. The $0^+$ strength functions

Figure 1 shows the $0^+$ IS and IV smoothed strength functions for $^{132}$Sn calculated with 100 Arnoldi iterations compared with the standard RPA results from Ref. [17]. Agreement between the strength functions is excellent. The four panels of Fig. 2 show the convergence of the smoothed strength functions of Fig. 1 as the number of Arnoldi iterations increases. The panels show differences of the strength functions calculated at 20 iteration intervals. The convergence is quite satisfactory after 100–120 iterations.

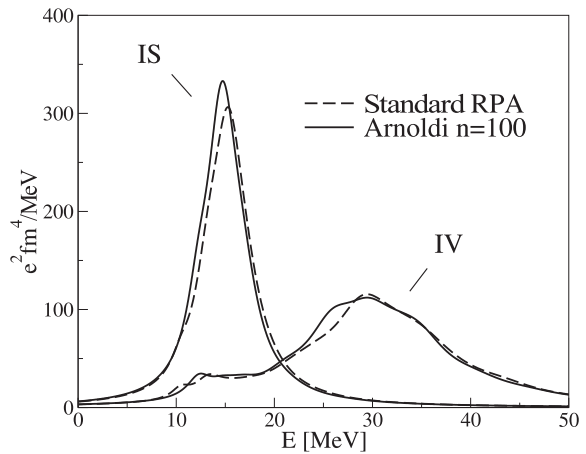

FIG. 1: The $0^+$ strength functions in $^{132}$Sn calculated by using 25 HO shells and 100 Arnoldi iterations for the SLy4 functional (solid lines), compared with the standard RPA calculation of Ref. [17] obtained for the SkM* functional (dashed lines).

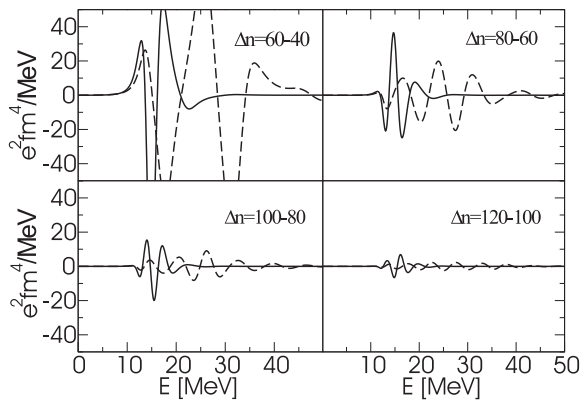

FIG. 2: Convergence of the $^{132}$Sn $0^+$ strength functions of Fig. 1. Solid lines are for the IS and dashed lines are for the IV strength functions. Each panel shows the difference of two strength functions, one with $n$ iterations and the other calculated with $n-20$ iterations.

### 2. The $2^+$ strength functions

Figures 3 and 4 show similar results as Figs. 1 and 2, but for the $2^+$ strength functions in $^{132}$Sn. As for the $0^+$ case, the IS and IV strength functions from the Arnoldi iteration agree very well with the strength functions of Ref. [17]. The convergence of strength functions is as fast as for $0^+$; after 120 iterations the smoothed strength functions change only by about 5%. We thus have to make only 120 iterations to calculate reasonably accurate $2^+$ strength functions for the RPA problem whose dimension is $D = 1020$. The large double spikes observed in Fig. 4 below 10 MeV are due to the lowest RPA phonons, which by the smoothing procedure acquire $\simeq$100-keV widths and move slightly down in excitation energy.

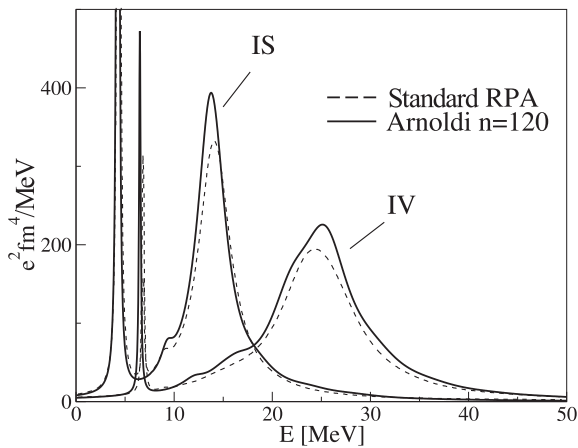

FIG. 3: Similar to Fig. 1 but for the $2^+$ strength functions. All results were calculated for the SLy4 functional.

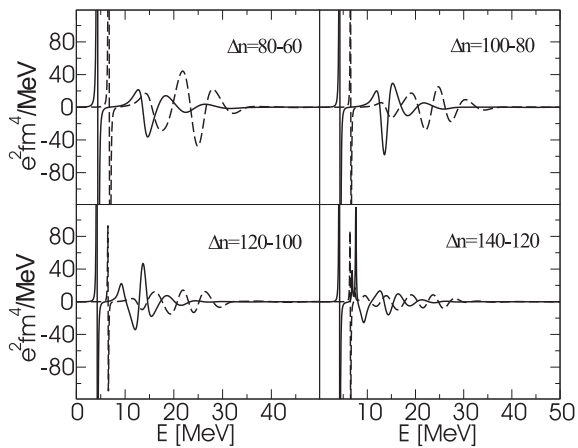

FIG. 4: Similar to Fig. 2 but for the $2^+$ strength functions.

### 3. The $1^-$ strength functions

Figure 5 compares the $1^-$ strength functions of our iterative method (solid line) with the strength functions from Ref. [17] (dashed line). The solid line shows the IS strength calculated when the generated Arnoldi basis is orthogonalized against spurious mode during iteration and dotted line corresponds to similar iterative calculation without orthogonalization. The low-lying state at 0.72 MeV, which has a large overlap with the spurious IS $1^-$ mode disappears when the orthogonalization method of Eqs. (22)–(24) is used. Also for the $1^-$ strength function, 100–120 Arnoldi iterations were needed to produce reasonably accurate results, see Fig. 6.

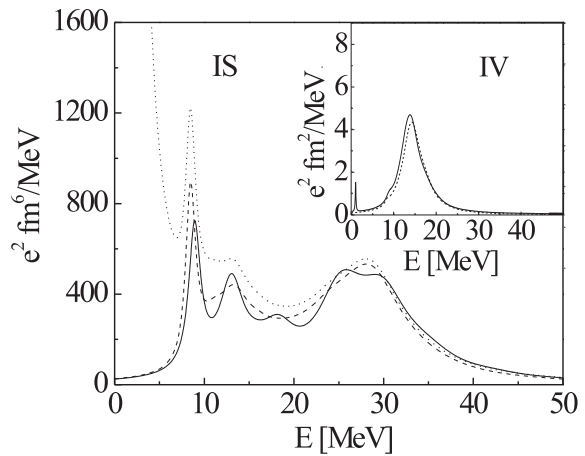

FIG. 5: Main panel: the $1^-$ strength functions in $^{132}$Sn, calculated using 100 Arnoldi iterations and with spurious IS mode removed (solid line), and results of the standard RPA from Ref. [17] (dashed line). Dotted line shows results of 140 Arnoldi iterations without orthogonalization against the spurious IS mode. Inset: same as in the main panel, but for the IV strength functions. All results were calculated for the SLy4 functional.

When no orthogonalization is made against the spurious mode, the obtained excitations contain small components of the spurious mode. This affects the physical part of the IS strength distribution, especially around 20–30 MeV. The standard RPA strength function of Ref. [17] has not been corrected for the spuriosity but only the strength of the lowest-lying state that has a large overlap with the spurious IS mode has been omitted. At 20–30 MeV, this strength agrees well with our uncorrected strength.

The orthogonalization method improves the convergence of the strength function, because now the 8.3 MeV $1^-$ excitation is lowest in energy and thus converges first. Without orthogonalization against the spurious mode, we need 140 iterations instead of 100 to get acceptably converged strength function.

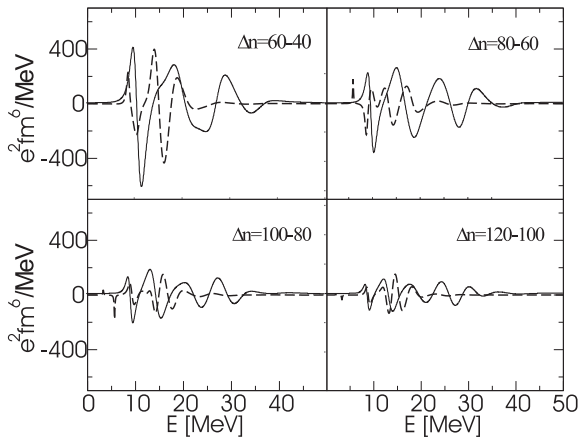

FIG. 6: Similar to Fig. 4 but for the $1^-$ strength functions. The IV strength-function differences were multiplied by the factor of $200\,\mathrm{fm}^4$.

## B. Convergence in function of the number of HO shells

In Section V A, we showed our strength functions calculated with 25 HO shells. This was found to be satisfactory, and using more shells did not appreciably change the obtained strength functions. The only effect of using more oscillator shells was that we needed to use slightly more Arnoldi iterations to produce well converged results. In the case of 40 shells, about twenty more iterations were needed, compared to calculations made with 25 shells. In Figs. 7, 8, and 9, we show the convergence of strength functions for the $0^+$, $2^+$, and $1^-$ modes, respectively. Each panel shows the difference of two strength functions obtained in the intervals of $\Delta N_0 = 4$ HO shells, between $N_0$ of 22 and 38.

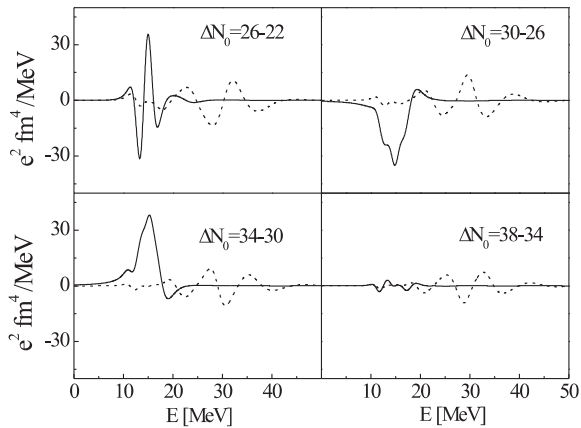

FIG. 7: Similar as Fig. 2 but for the convergence of the $0^+$ strength functions as a function of the number of HO shells $N_0$.

These plots overstress the variations of strength functions in the sense that slight shifts of peaks create the oscillating patterns in the difference plots. To illustrate this point, in Fig. 10 we show the $1^-$ strength functions calculated for $N_0 = 22$, 26, 30, 34, and 38 HO shells. Poor convergence of the IS surface mode creates some uncertainty in the position and width of the high-energy bump. Larger bases should probably be used if converged results for this particular mode were required.

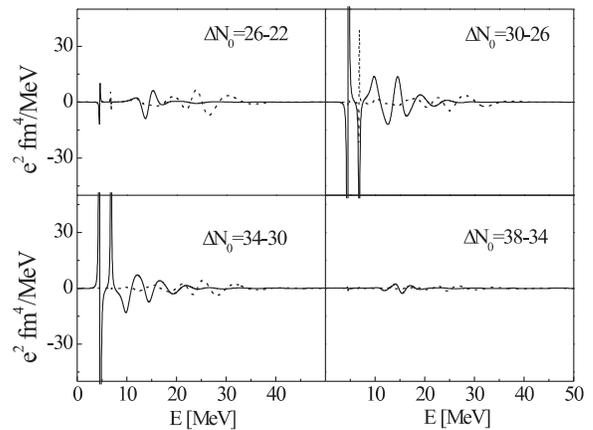

FIG. 8: Similar as Fig. 4 but for the convergence of the $2^+$ strength functions as a function of the number of HO shells $N_0$.

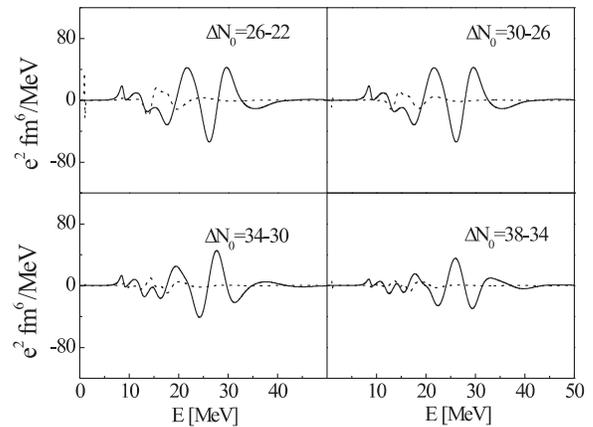

FIG. 9: Similar as Fig. 6 but for the convergence of the $1^-$ strength functions as a function of the number of HO shells $N_0$.

As noted in Ref. [5], well before the maximum number of iterations (equal to the RPA dimension $D$) is reached, the iteratively generated RPA matrix in the Krylov space can become singular. In that case, the stabilized iteration method of Eqs. (12)–(19) still protects us from obtaining complex RPA eigenvalues, but the condition number of the Krylov-space RPA matrix approaches infinity, because one or more of its eigenvalues collapse nearly to zero.

In the standard method, the RPA matrix is calculated by using the bare p-h basis states. In our method, we instead start from the pivot vector of Eq. (20) and the Arnoldi iteration then produces the rest of our basis vectors composing the Krylov subspace. This subspace is spanned by the eigenstates of the RPA matrix which has an overlap with the pivot vector. Thus, in general, the Arnoldi iterations can only be continued until this subspace is exhausted in which case the condition number goes to infinity. However, with finite numerical precision this maximum limit of Arnoldi iterations is further reduced.

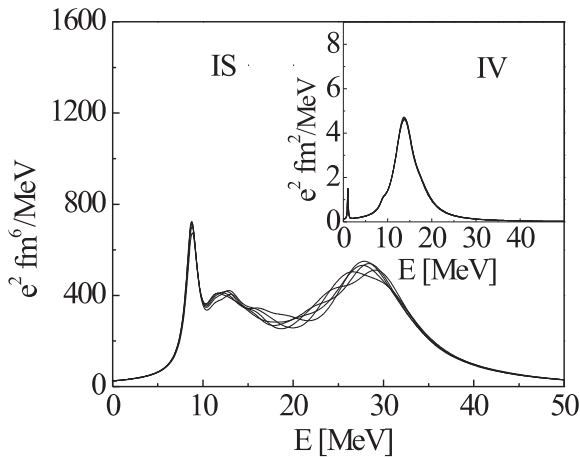

FIG. 10: Similar as Fig. 5 but for the $1^-$ strength functions calculated for the numbers of HO shells $N_0 = 22, 26, 30, 34,$ and 38

TABLE I: Spherical RPA and QRPA dimensions $D$ as functions of the number of HO shells $N_0$.

|       | RPA |     |      | QRPA |       |       |
|-------|-----|-----|------|------|-------|-------|
| $N_0$ | $0^+$ | $1^-$ | $2^+$ | $0^+$ | $1^-$ | $2^+$ |
| 10    | 70  | 195 | 261  | 390  | 1040  | 1510  |
| 20    | 205 | 555 | 766  | 2880 | 8180  | 12720 |
| 25    | 273 | 734 | 1020 | 5538 | 15912 | 25088 |
| 30    | 340 | 915 | 1271 | 9470 | 27420 | 43630 |

In a typical iteration, during the first few iterations the condition number of the Krylov-space RPA matrix fluctuates, then approaches a stable plateau, and finally suddenly goes toward infinity. When that happens, the iteration must be stopped and one must backtrack to the iteration where the condition number was still acceptable. Therefore, the number of Arnoldi iterations can depend on the size of the HO basis, and the results presented in this section correspond to the numbers of iterations fixed according to this prescription.

## VI. SCALING OF ITERATIVE SOLUTION METHOD

We illustrate the benefits of the iterative solution of the RPA or QRPA equations over the traditional method by comparing how the numerical work increases in the iterative method as the HO basis is increased.

As can be seen in Table I, the RPA dimensions $D$ for doubly magic spherical nuclei increase almost linearly with the number of oscillator shells $N_0$. This is easy to understand, because in this case, only the number of particle states increases and the number of hole states always stays constant. Therefore the time to solve the full RPA eigenproblem in this case scales approximately as $N_0^3$. In the spherical QRPA, the dimensions scale roughly between $N_0^2$ and $N_0^3$, and the full QRPA scales approximately as $N_0^6$ or $N_0^9$. The physically interesting and computationally challenging calculations are for deformed nuclei with pairing, and we should therefore compare the $N_0$ scaling of iterative and standard QRPA diagonalizations.

In the case of all symmetries of the mean field being broken, the QRPA dimension $D$ is:

$$D = \frac{1}{9}[(N_0 + 1)(N_0 + 2)(N_0 + 3)]^2 . \qquad (25)$$

This dimension increases very steeply ($N_0^6$) as the number of HO shells is increased. For $N_0 = 14$, the QRPA dimension is $D = 1849600$, for example. The corresponding standard QRPA solution scales as $N_0^{18}$ and is thus untractable. Therefore, for the QRPA calculations in deformed nuclei, we must truncate the single-particle space. The best method is to use the two-basis method [21], by which one solves the HFB equations in the basis generated by the HF part of the HFB matrix, and truncate the basis using a cutoff on the obtained pseudo-HF single-particle energies. But even then, the QRPA calculations scale as the sixth power of the number of useful single-particle states, and are thus prohibitively difficult.

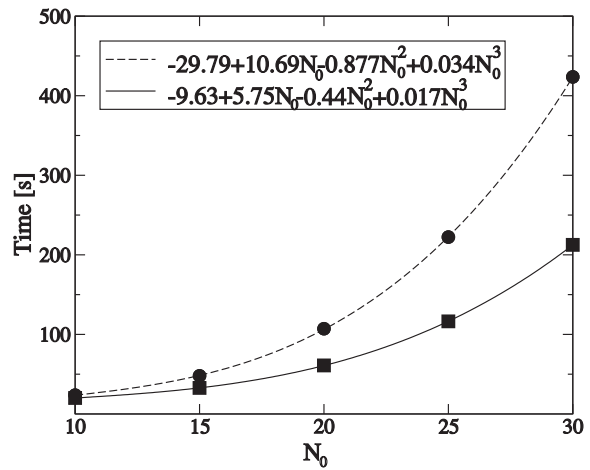

FIG. 11: Times to calculate 100 Arnoldi iterations for the spherical QRPA method applied to $^{132}$Sn as functions of $N_0$. Squares and circles show results for the $1^-$ and $2^+$ modes, respectively, and lines show cubic fits.

In order to illustrate the scaling properties of the iterative QRPA method, we calculated the corresponding matrix-vector products with our developmental QRPA code, where the pairing has been set to zero. The RPA fields $\tilde{h}$ only depend on the normal RPA density matrix and the calculation of pairing part of the QRPA matrix-vector product is very fast due to a small number of the pairing coupling constants and the simple density dependence of typical pairing EDF. Therefore the time to calculate the pairing part is negligible. The only significant increase of running time in spherical QRPA compared to RPA comes from the need to handle higher-dimensional basis vectors in the calculation of various overlaps and vector additions during the Arnoldi iteration. Therefore, as we keep all particle-particle and hole-hole RPA amplitudes in the calculation, but set them to zero, our timing results accurately reflect the timing of the spherical QRPA calculation.

In Fig. 11, we show the scaling properties of the spherical QRPA calculation with iterative Arnoldi method. It is clear that the scaling of our iterative method is as $N_0^3$, that is, it is linear with respect to the QRPA dimension $D$. Of course, the prefactor itself is linearly proportional to the number of Arnoldi iterations. However, as discussed in the previous section, the Arnoldi iteration method cannot in practice go full dimension before the generated Krylov-space matrices become singular. As long as we are satisfied with a few hundred iterations at most, the iterative method gives us a vast speed improvement. In the full RPA or QRPA diagonalization, the calculation and storage of a very large dense RPA or QRPA matrices also takes a considerable additional time – a step that the iterative method avoids completely.

In addition to the moment-method based iteration, which is ideal for strength functions, the iterative method can also be modified to be suitable for different kinds of other calculations. If we are interested in a number of very well converged lowest RPA eigenmodes, restarted Arnoldi methods [22] can be used. These methods use more iterations than basis states, i.e., after a maximum number $d$ of basis vectors is generated, new approximations for the wanted $d' < d$ eigenmodes are calculated, and iteration is then continued to generate new improved basis states from $d'+1$ to $d$ again. The restarting can be made as many times as needed to produce wanted number of well converged lowest excitations.

Methods such as Arnoldi or Lanczos produce convergence at the extreme ends of the excitation energy spectrum. If eigenmodes away from the extremes are looked for, shift and invert methods [23, 24] can be used. These methods allow iterative methods to be used to find RPA eigenmodes anywhere inside the RPA excitation spectrum.

## VII. SUMMARY AND CONCLUSIONS

We have presented a method to calculate accurate RPA response functions by using the iterative Arnoldi diagonalization related to the sum-rule conserving Lanczos method of Ref. [5]. We used strictly the same EDF for the ground state calculation and RPA excitations. We have showed how the Arnoldi method must be stabilized in order to apply it reliably to the RPA eigenvalue problem. The resulting electromagnetic strength functions are in good agreement with the standard RPA results and are obtained with numerical effort smaller by orders of magnitude.

Our method closely resembles the FAM of Nakatsukasa et al. [6, 7], except that our iterative method is different and that we use the HO basis instead of the mesh in coordinate space. The FAM and our method both allow the existing EDF mean-field codes to be used for the calculation of the RPA or QRPA matrix-vector products. With minor modifications, mostly pertaining to the full implementation of the time-odd mean fields, these codes can easily be extended to RPA/QRPA. In particular, our future implementation of the deformed QRPA solution will be based on the code HFODD [25].

We also implemented the method to remove components of the spurious RPA modes from the calculated strength functions that keeps the physical excitations exactly orthogonal against the spurious excitations in any finite model space.

The smaller numerical effort of the iterative Arnoldi method, and the fact that in this method one does not have to calculate and store the RPA or QRPA matrices, allows our method to be applied to the calculation of electromagnetic and beta decay strengths and strength functions for deformed heavy nuclei. Work to extend our formalism and codes to deformed superfluid nuclei is in progress.

### Acknowledgments

We are thankful to J. Terasaki for providing us with numerical values of the strength functions calculated in Ref. [17]. This work was supported by the Academy of Finland and University of Jyväskylä within the FIDIPRO program and by the Polish Ministry of Science and Higher Education under Contract No. N N 202 328234.

[1] S. Tretiak, C.M. Isborn, A.M.N. Niklasson, and M. Challacombe, J. Chem. Phys. **130**, 054111 (2009).
[2] M.W. Schmidt, K.K. Baldridge, J.A. Boatz, S.T. Elbert, M.S. Gordon, J.H. Jensen, S. Koseki, N. Matsunaga, K.A. Nguyen, S. Su, T.L. Windus, M. Dupuis, and J.A. Montgomery, Jr, J. Comput. Chem. **14**, 1347 (1993).
[3] http://www.msg.ameslab.gov/gamess/gamess.html.
[4] P.-G. Reinhard, Ann. Physik **1**, 632 (1992).
[5] C.W. Johnson, G.F. Bertsch, and W.D. Hazelton, Comput. Phys. Commun. **120**, 155 (1999).
[6] T. Nakatsukasa, T. Inakura, and K. Yabana, Phys. Rev. C **76**, 024318 (2007).

[7] T. Inakura, T. Nakatsukasa, and K. Yabana, Phys. Rev. C **80**, 044301 (2009).
[8] S. Péru and H. Goutte, Phys. Rev. C **77**, 044313 (2008).
[9] K. Yoshida and Nguyen Van Giai, Phys. Rev. C **78**, 064316 (2008).
[10] J. Terasaki *et al.*, unpublished.
[11] C. Losa *et al.*, unpublished.
[12] P. Ring and P. Schuck, *The Nuclear Many-Body Problem* (Springer-Verlag, Berlin, 1980).
[13] J.P. Blaizot and G. Ripka, *Quantum theory of finite systems*, MIT Press, Cambridge Mass., 1986.
[14] B.K. Agrawal and S. Shlomo, Phys. Rev. C **70**, 014308 (2004).
[15] Tapas Sil, S. Shlomo, B.K. Agrawal, and P.-G. Reinhard, Phys. Rev. C **73**, 034316 (2006).
[16] J. Terasaki, J. Engel, M. Bender, J. Dobaczewski, W. Nazarewicz, and M. Stoitsov, Phys. Rev. C **71**, 034310 (2005).
[17] J. Terasaki and J. Engel, Phys. Rev. C **74**, 044301 (2006).
[18] Y. Saad, *Iterative methods for sparse linear systems (2nd ed.)* (SIAM, 2003).
[19] B.G. Carlsson, J. Dobaczewski, J. Toivanen, and P. Veselý, to be published in Computer Physics Communication.
[20] E. Chabanat, P. Bonche, P. Haensel, J. Meyer, and R. Schaeffer, Nucl. Phys. **A627** (1997) 710.
[21] B. Gall, P. Bonche, J. Dobaczewski, H. Flocard, and P.-H. Heenen, Z. Phys. **A348**, 183 (1994).
[22] R.B. Lehoucq and D.C. Sorensen, SIAM J. Matrix Anal. Appl. **17**, 789 (1996).
[23] Y. Saad, *Numerical methods for large eigenvalue problems, Algorithms and architectures for advanced scientific computing* (Manchester University Press, Manchester, 1992).
[24] Zhongxiao Jia and Yong Zhang, Computers Math. Applic. **44**, 1117 (2002).
[25] J. Dobaczewski *et al.*, Comput. Phys. Commun. **102**, 166 (1997); **102**, 183 (1997); **131**, 164 (2000); **158**, 158 (2004); **167**, 214 (2005); **180**, 2361 (2009).